\documentclass[aps,prb,groupedaddress,showpacs,twocolumn,superscriptaddress]{revtex4}
\usepackage{graphicx}
\usepackage{amsmath}
\usepackage{amssymb}
\usepackage{bbm}
\usepackage{xspace}
\usepackage{color}
\usepackage{footnote}
\usepackage{comment}
\usepackage{placeins}
\usepackage{hyperref}
\usepackage{tcolorbox}
\usepackage{dsfont}
\usepackage{xcolor}
\usepackage[normalem]{ulem}

\newcommand{\etal}[0]{\textit{et al.}}

\newcommand{\fs}{{\tilde{f}}}

\def\nr2#1{\mathinner{\lvert\lvert{#1}\rvert\rvert^2}}

\definecolor{orange}{RGB}{252,77,6}
\definecolor{brown}{RGB}{200,127,50}
\definecolor{green}{RGB}{0,200,100}

\begin{document}

\title{Non-equilibrium Green's functions and their relation to the negative differential conductance in the interacting resonant level model}

\author{Max E. Sorantin}
\email[]{sorantin@tugraz.at}
\affiliation{Institute of Theoretical and Computational Physics, Graz University of Technology, 8010 Graz, Austria}
\author{Roman Lucrezi}
\affiliation{Institute of Theoretical and Computational Physics, Graz University of Technology, 8010 Graz, Austria}
\author{Wolfgang von der Linden}
\affiliation{Institute of Theoretical and Computational Physics, Graz University of Technology, 8010 Graz, Austria}
\author{Enrico Arrigoni}
\email[]{arrigoni@tugraz.at}
\affiliation{Institute of Theoretical and Computational Physics, Graz University of Technology, 8010 Graz, Austria}

\date{\today}

\begin{abstract}
We evaluate the non-equilibrium single particle Green's functions in the steady state of the interacting resonant level model (IRLM) under the effect of an applied bias voltage. Employing the so-called auxiliary master equation approach, we present accurate nonperturbative results for the non-equilibrium spectral and effective distribution functions, as well as for the current-voltage characteristics. We find a drastic change of these spectral properties between the regimes of low and high bias voltages and discuss the relation of these changes to the negative
 differential conductance (NDC), a prominent feature in the non-equilibrium IRLM. 
The anomalous evolution of the effective distribution function next to the impurity shown by our calculations
suggests a mechanism whereby the impurity gets effectively decoupled from the leads
at voltages where the NDC sets in, in agreement with previous 
renormalization group approaches.
This scenario is qualitatively confirmed by a Hartree-Fock treatment of the model. 
\end{abstract}


\maketitle

\section{Introduction}\label{sec:Introduction}

Transport through 
nanodevices
 such as molecular junctions or quantum dots has become of great interest in the past due to the potential application of these systems as new type of electronic components\cite{ra.13,ko.sm.14}. Generically, the working principle of such components is entailed in
their current/voltage (I/V) characteristic.
In some situations this
can display non-monotonic behavior, usually referred to as negative differential conductance (NDC), a peculiar effect that is intriguing by itself but also most useful in potential applications\cite{ce.re.1999,ce.wa.2000,kr.ir.02,xu.su.1999,hu.wu.14}. Therefore, a thorough understanding of the NDC is highly desirable.

A prototypical model exhibiting a NDC is the so-called interacting resonant level model (IRLM), a simplistic model featuring a two level quantum dot connected to leads used to study the interplay of quantum fluctuations and electronic correlations in the setting of quantum impurity problems. Introduced by Vigman and 
Finkelstein\cite{wi.fi.78}
in the (equilibrium-) context of the Kondo-problem, the IRLM in non-equilibrium has received increasing attention over the last decade after the discovery of an analytic expression for the I/V characteristic\cite{bo.sa.08} in the so-called scaling regime and for a special value of the interaction, referred to as the self-dual point of the IRLM.

Previous works on the IRLM in non-equilibrium extended the analytic treatment of the self-dual point\cite{go.sc.18u}, considering also higher order statistics of charge transport\cite{br.bo.10,ca.ba.11}, and provided further validation by numerical treatment of increasing accuracy\cite{bi.mi.17}. Away from the self-dual point, Perfetto \etal\cite{pe.st.12} studied the transport properties of the IRLM employing non-equilibrium Green's functions (NEGF) focusing on the effect of long-range interactions. In addition, a perturbative treatment within NEGF\cite{vi.sc.14} as well as
renormalization group (RG) approaches\cite{an.me.10,ke.me.13} 
,valid for weak interactions,
provide further insight for small interactions. In particular, it is found that the NDC within RG arises due to a renormalization of the hopping rate into the leads which gets suppressed for higher voltages\cite{bo.vl.07,ca.an.10,ka.pl.10}. In contrast, less is known about the physical mechanism of the NDC at the self-dual point,
 i.e., for intermediate values of the interaction.
Related to the NDC, but also very interesting in itself, is the spectral function of the IRLM which in equilibrium was numerically studied by Braun and Schmitteckert\cite{br.sc.14}
but, to our knowledge, has not been considered so far in a non-equilibrium situation
within a non-perturbative treatment.

In this work, 
we evaluate NEGF of the IRLM 
in order to investigate 
their connection with the
NDC  and how the spectral and effective distribution functions evolve in terms of the bias voltage.
Our results are obtained within
the so-called auxiliary master equation approach (AMEA) - a numerical method to treat non-equilibrium quantum impurity problems and evaluate their NEGF with considerable accuracy.
For simplicity, our calculations refer to the self-dual point, but can be readily carried out for other values of the interaction.
Finally, we complement our discussion of the AMEA results with an Hartree-Fock (HF) treatment in order to help with the interpretation.

We find that in the regime of the NDC, the spectral function 
evolves from a peak at finite frequencies into
a dominant central peak and that the NDC can be traced back to the behavior of the effective distribution functions on the first lead sites.
We interpret this behavior as an effective decoupling of the impurity from the leads, which is 
confirmed be the HF calculations.

\section{Model and Method}\label{sec:ModelMethod}

\subsection{Model}\label{subsec:Model}

\begin{figure}
 \includegraphics[width=0.9\columnwidth]{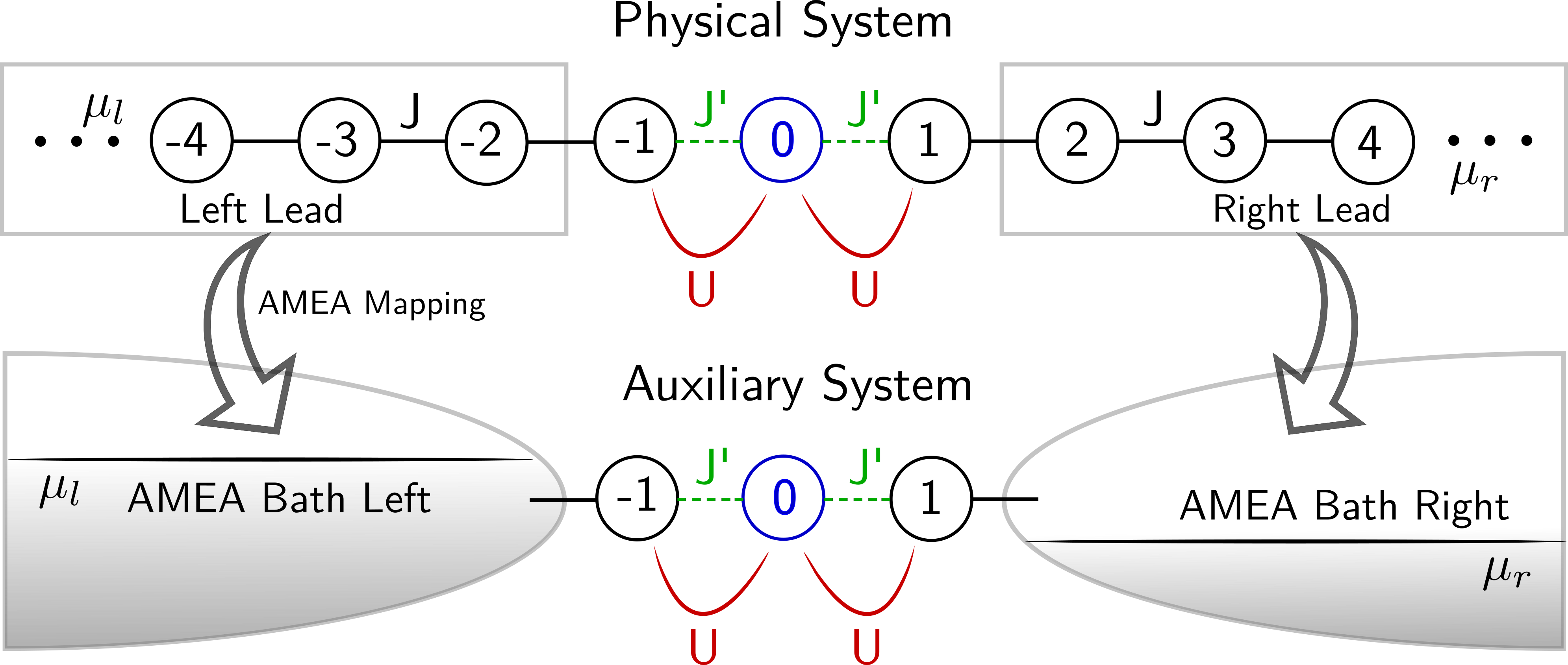}
 \caption{A sketch of the IRLM as lattice model and its mapping to the auxiliary open quantum system used within AMEA.
}
\label{fig:IRLMsketch}
\end{figure}

The IRLM is a well-known  impurity model of spinless fermions. It features an impurity site connected to two semi-infinite tight-binding chains together with a density-density interaction term coupling the impurity site to the neighboring chain sites, see Fig.~\ref{fig:IRLMsketch}. The Hamiltonian is defined as,
\begin{align}
 \notag H_{\text{IRLM}} =& H_{\text{L}} + H_{\text{R}} + H_{\text{dot}}\\
 \notag H_{\text{L}} =& -J\sum_{r = -\infty}^{-2} c_{r}^{\dagger}c_{r+1} + h.c.\\
 \notag H_{\text{R}} =& -J\sum_{r = 1}^{+\infty} c_{r}^{\dagger}c_{r+1} + h.c.\\
 \notag H_{\text{dot}} =& -J' \sum_{r=-1}^{r=0}c_{r}^{\dagger}c_{r+1} + h.c.\\
 &+U\sum_{r=\pm 1} \left( c_{r}^{\dagger}c_{r} - \frac{1}{2}\right)\left(c_{0}^{\dagger}c_{0} - \frac{1}{2}\right)
 \label{eq:Hamiltonian}
\end{align}
where $c_{r}^{\dagger}/c_{r}$ denote the fermionic creation/annihilation operators at site $r$. Here, $H_{\text{L/R}}$ describe the semi-infinite tight-binding chains and $H_{\text{dot}}$ introduces the hopping to the impurity as well as the interaction term. A non-equilibrium steady state situation is induced in the system via an applied bias voltage $V$ simulated by shifting the chemical potentials of the leads symmetrically, that is $\mu_l=-\mu_r=\frac{V}{2}$. We use $J$ as unit of energy and work in units where $\hbar=e=k_B=1$.

\subsubsection{The continuum limit and the scaling regime of the IRLM}

\label{subsubsec:IRLM_scalingregime}
 Here, we want to summarize some well known facts about the IRLM in the so-called scaling regime, which are important for the present work. A nice overview in the non-equilibrium context can be found in the recent works \cite{bi.mi.17,ca.ba.11,wi.fi.78,bo.sa.08} and references therein.\\
 When the bandwidth of the leads, $W=4J$, is the dominant energy scale in the system the lattice model, Eq.~\ref{eq:Hamiltonian}, becomes equivalent to its continuum limit\footnote{also referred to as wide-band limit}, allowing for a field theoretic description. In this scaling regime of the IRLM, the physics becomes universal with the emergence of a Kondo energy scale $T_B\sim (J')^\frac{4}{3}$.
 The constant of proportionality is the lattice regularization of the corresponding field theory relating results from the continuum limit to the lattice model.\\
 The continuum model can be solved analytically for the special value of the interaction $U_{\text{c}}^*=\pi$, which corresponds to $U_{\text{lat}}^*\cong2$ in the lattice model, where the IRLM exhibits a certain self-duality. Most notably, there is a closed form expression for the steady state current at $T=0$
\begin{equation}
 I(V) = \frac{V}{2\pi} \hspace{1pt} _{2}F_{3}\left[\left\{ \frac{1}{4},\frac{3}{4},1 \right\}, \left\{ \frac{5}{6},\frac{7}{6} \right\}; -\left(\frac{V}{V_c}\right)^6 \right]
 \label{eq:Currentexpression}
\end{equation}
with $V_c=c(J')^{\frac{4}{3}}$, where $c\approx 3.2$
\footnote{In more detail, $V_c = \frac{\sqrt{3}}{4^{2/3}} \frac{4\sqrt{\pi}\Gamma (2/3)}{\Gamma (1/6)} T_B$ and $T_B = \tilde{c}(J')^{\frac{4}{3}}$ with $\tilde{c}\approx 2.7$ from \cite{bo.sa.08}.} and $_{2}F_{3}(a,b;z)$ is the generalized hypergeometric function. From Eq.~\ref{eq:Currentexpression}, it immediately follows that $I/V = f(V/V_c)$ depends only on the rescaled voltage and thus has a universal form set by the energy scale $T_B$. As is best seen by expanding the hypergeometric function up to leading order
\begin{equation}
 I(V\ll V_c) \approx \frac{V}{2\pi}\left[ 1 - \frac{24}{170}\left(\frac{V}{V_c}\right)^6 + O\left(\frac{V}{V_c}\right)^{12}\right]
 \label{eq:Currentsmallvoltages}
\end{equation}
the current is linear for small voltages $V<V_c$. The most prominent feature of the current arises for $V>V_c$ where the model exhibits a negative differential conductance, see Fig.~\ref{fig:Current}.\\

\subsection{Method}\label{subsec:Method}
 In this work, we use the Auxiliary Master Equation Approach (AMEA) \cite{ar.kn.13,do.nu.14,do.so.17} to investigate the IRLM under the influence of an applied bias voltage. AMEA is a method to treat non-equilibrium correlated impurity problems which is particularly efficient to target the steady state. 
It is based upon mapping the noninteracting bath onto an auxiliary open quantum system whose dynamics is described by the Lindblad equation. 
This mapping becomes exponentially accurate by increasing the number of sites in this auxiliary system.
This open quantum systems effectively mimics a system with infinite volume, so that one can reliably reach the steady state. Correlation functions are then obtained by time evolution of the many-body density matrix starting from the steady state.\\
The dynamics of the auxiliary open quantum system can be solved  numerically exact by available approaches. Here,  we employ the  so-called Stochastic-Wavefunctions\cite{da.ca.92,br.ka.97,br.ka.98}, whose application to AMEA is presented in Ref.~\cite{so.fu.18u}.
 Within the mapping, the central interacting region  $|r|\leq 1$ described by $H_{\text{dot}}$ (cf. Fig.~\ref{fig:IRLMsketch}) remains unchanged~\footnote{We note that the mapping needs to be performed for each bias voltage separately but is independent of the parameters in $H_{\text{dot}}$, i.e. $J'$ and $U$.}
 In total the auxiliary open quantum system, thus, consists of $L=3 + 2N_B$ sites, where $N_B$ denotes the number of auxiliary dissipative bath levels used to replace the left(right) semi-infinite leads. For details, we refer to previous publications~\cite{so.fu.18u,ar.kn.13,do.nu.14,do.so.17}.

\subsubsection{Steady-state Current}\label{subsubsec:SSCurrent}
The  current 
$I_{r,r+1}$
across a bond connecting site $r$ and $r+1$, which is clearly independent of $r$ in the steady state,
can be expressed within the Keldysh Green's function (GF) formalism as~\cite{ha.ja}
\begin{align}
\notag I_{r,r+1} &= t_{r,r+1}^2 \int \frac{d\omega}{2\pi} \text{Re}j(\omega)\\
 j(\omega) &= 
 G_{rr}^R(\omega)g_{r+1,r+1}^K(\omega) + G_{rr}^K(\omega)g_{r+1,r+1}^A(\omega) \;
 \label{eq:CurrentfromGF}
\end{align}
provided the interaction self-energy is zero across the bond.
Here, a capital $G_{r,r}(\omega)$ denotes the local  GF of the full system, while the  
  lower case $ g_{r,r}(\omega)$ is the one when the system is disconnected at the bond connecting the sites $r$ and $r+1$. 
A convenient choice is the bond from one noninteracting bath to the interacting region, i.e. $r=-2$ to $r=-1$. 

In equilibrium, $V=0$, the Keldysh and retarded GF are not independent and connected by the fluctuation-dissipation theorem, which for the GF's appearing in Eq.\eqref{eq:CurrentfromGF} reads
\begin{align*}
\text{Im}\; G^K_{rr'}(\omega) &= 2(1-2f(\omega))\text{Im}\;G^R_{rr'}(\omega) \, ,
\end{align*}
where $f(\omega)$ denotes the fermi-dirac distribution function.
In analogy, one can define
an effective local non-equilibrium energy distribution function $\fs_r(\omega)$ 
 via the relation (cf. also e.g.~\cite{ts.ok.09,mu.ts.17,ha.ji.13,ca.ko.12,ca.we.13})
\begin{align}\label{eq:neqFD}
\text{Im}\; G^K_{rr}(\omega) &= 2(1-2\fs_{r}(\omega))\text{Im}\;G^R_{rr}(\omega) \, .
\end{align}

With Eq.\eqref{eq:neqFD},
we can express the current 
 from the left lead into the central region as\footnote{Which reduces to the well known Landauer formula for $U=0$ involving only the difference of the left/right Fermi-functions. To see this consider Eq.\eqref{eq:HF_s}.}
\begin{align}
 \notag I_L(V)\equiv I_{-2,-1} = 2\pi (J)^2 \int d\omega & A_{-1}(\omega;V)A_{TB}(\omega)\\
 \times&\left[ f_L(\omega;V) - \fs_{-1}(\omega;V) \right]\, ,
 \label{eq:il}
\end{align}
where $A_r\equiv-\frac{1}{\pi} \text{Im}G_{rr}^R$ is the local density of states, $A_{TB}(\omega)$ denotes the DOS of the disconnected left lead, that is the DOS of a semi-infinite tight binding chain, and $f_L$ is the fermi-function of the left lead.
Here, for convenience,
 we have indicated any possible dependence on the bias-voltage.
In Eq.\eqref{eq:il}, the frequency integrand contains
 the difference between the effective distribution function at the first correlated site $r=-1$ and the one deep into the left lead 
weighted with the corresponding DOS.

\section{Results}\label{sec:Results}
\begin{figure}
 \includegraphics[width = 0.99 \columnwidth]{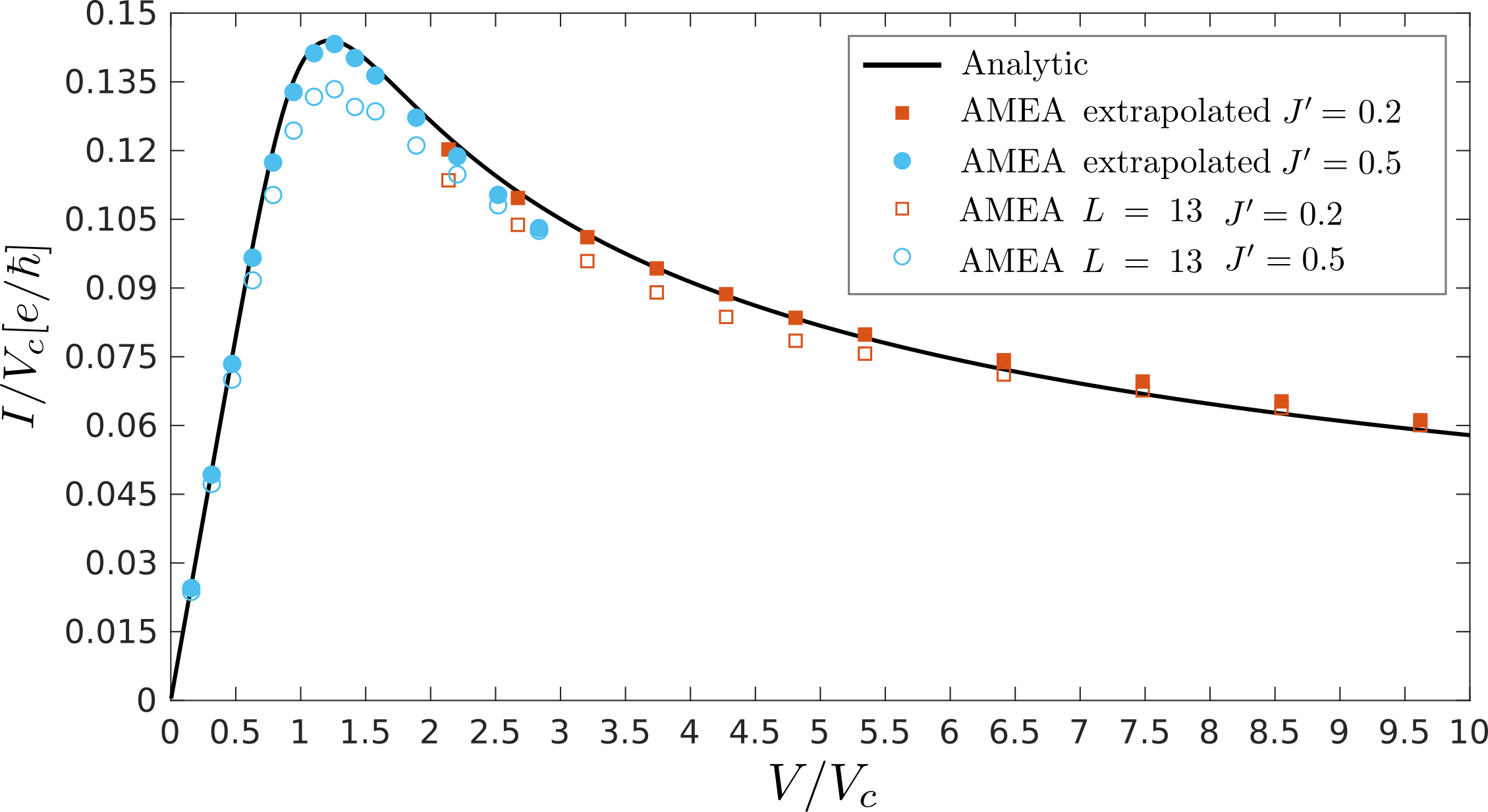}
 \caption{Current voltage characteristic of the IRLM for two different hybridization strengths $J'=0.5$(blue circles) and $J'=0.2$ (red squares). We display the analytic solution at $T=0$ (solid lines), the extrapolated, formally $L\rightarrow\infty$, AMEA data from ref\cite{so.fu.18u} (filled symbols) and the AMEA current in the $L=13$ system (open symbols).
 }
 \label{fig:Current}
\end{figure}

In this section, we present results for the non-equilibrium spectral properties of the IRLM. We are not aware of previous numerically accurate results for the non-equilibrium Green's function of this model from the literature.
We consider the self dual point $U=2$ and compute results for
two different values of the hybridization strength $J'=0.2$ and $J'=0.5$ at a finite temperature $T=0.025$.
The size of the auxiliary system, which controls the accuracy of the bath hybridisation function (see ~\cite{do.nu.14,do.so.17}) is fixed to $L=13$.
Both the steady state as well as the Green's functions are obtained by time evolution by stochastic wave functions, see  ~\cite{so.fu.18u} 
 for  technical details. 
In order to illustrate the accuracy of the approach, we first plot the
 steady state current as function of the bias voltage in
Fig.\ref{fig:Current}. Specifically we compare data from the present $L=13$ auxiliary-system calculation
with the ones of the more accurate approach of Ref.~\cite{so.fu.18u}, where the current is obtained via an extrapolation for values of $L$ up to $L=19$. The analytic solution of the continuum model at $T=0$ is also shown for comparison.
 In this paper we use smaller values of $L$ because a full Green's function calculation for $L=19$ would be computationally too expensive.
These results show that also $L=13$ provides quite accurate results\footnote{see also ref\cite{so.fu.18u} for a benchmark of the $L=13$ GF in equilibrium.} and in particular reproduces  
 the NDC.

\subsection{Spectral properties at the central impurity site}\label{subsec:A_imp0}
\begin{figure}
 \includegraphics[width=0.9\columnwidth]{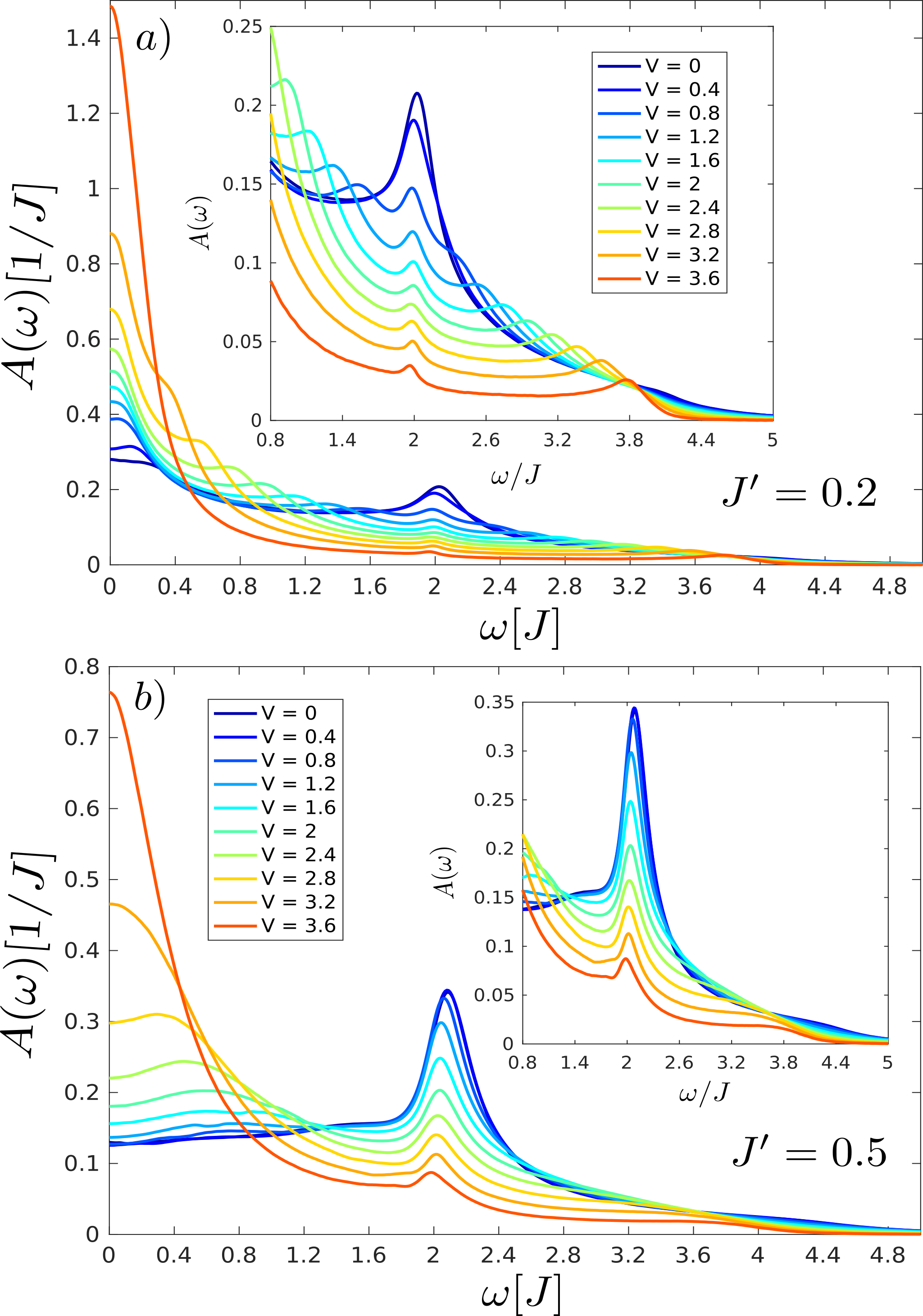}
 \caption{Local density of states at the impurity site, $r=0$, for different bias voltages. Upper panel, a) for $J'=0.2$, and lower panel b)  for $J'=0.5$, other parameters $T=0.025$ and $U=2$. The insets show a zoom around the peak at $\omega=2$ and its appearing satellite at $\omega=2\pm V/2$.
 \label{fig:DOSimp0}
}
\end{figure}

Fig.~\ref{fig:DOSimp0} shows the density of states at the impurity site, $r=0$, for different bias voltages~\footnote{We only show the result for positive frequencies since even at finite bias the IRLM is still particle-hole symmetric at the impurity site.}.The equilibrium ($V=0$) system is characterized by a pronounced peak at $\omega=2$. Upon increasing the bias voltage, 
the spectral weight is removed from the $\omega\approx 2$ in favor of a second peak at  zero frequency, which quickly becomes dominant  
for large bias voltages. 
At the same time, the equilibrium resonance develops sidebands at $\omega = 2 \pm V/2$.
 This effect is more pronounced for the case of low $J'=0.2$ since a stronger $J'$ broadens all peak features.
At large voltages, $V\gtrsim 3.2$ for $J'=0.2$, the left satellite merges with the central peak.

\begin{figure}
 \includegraphics[width=0.9\columnwidth]{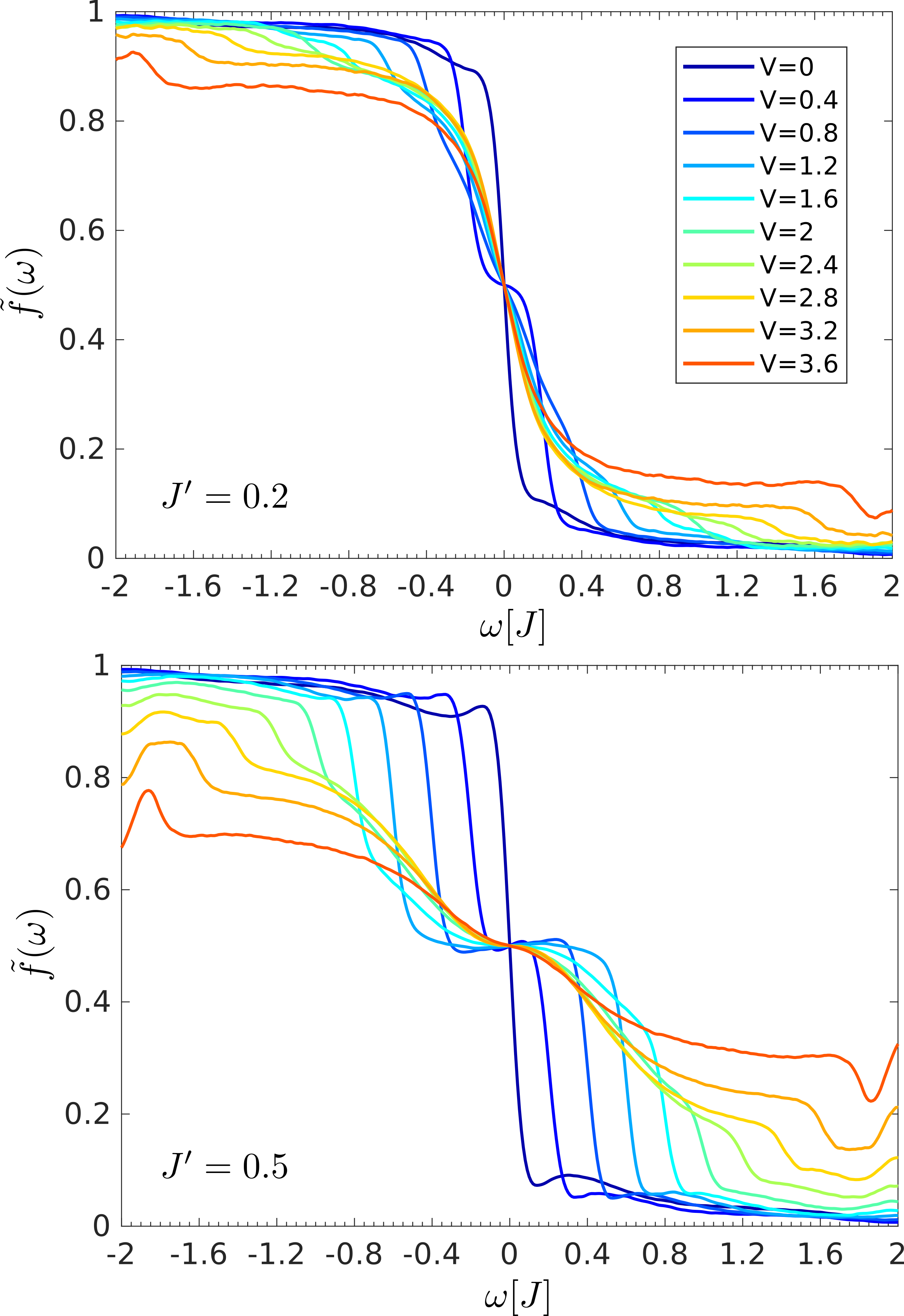}
 \caption{Distribution function at the central impurity site for different bias voltages. Parameters are as in Fig.~\ref{fig:DOSimp0}. The non-smoothness of the curves is due to the statistical error 
 of the SWF approach, which gets amplified for the effective distribution function as this is given by the ratio two Green's functions.
 \label{fig:Distimp0}
}
\end{figure}

Out of equilibrium the fermionic effective distribution function obviously deviates from the Fermi Dirac distribution and acquires an anomalous, position-dependent shape. 
In Fig.~\ref{fig:Distimp0}, we plot the effective distribution function, $\fs_{r}(\omega)$ defined in Eq.\eqref{eq:neqFD}, at the impurity site, $r=0$, for different bias voltages. We find, that the latter is 
dominated by a double Fermi-step, $2\fs_{r=0}(\omega) =f_L(\omega) + f_R(\omega) $, for small bias voltages and 
changes drastically its shape for bias voltages where the NDC sets in.

\subsection{Sites next to the impurity ($r=\pm 1$) and relation with the current integrands}\label{subsec:GF_minus1}
\begin{figure}
 \includegraphics[width=0.9\columnwidth]{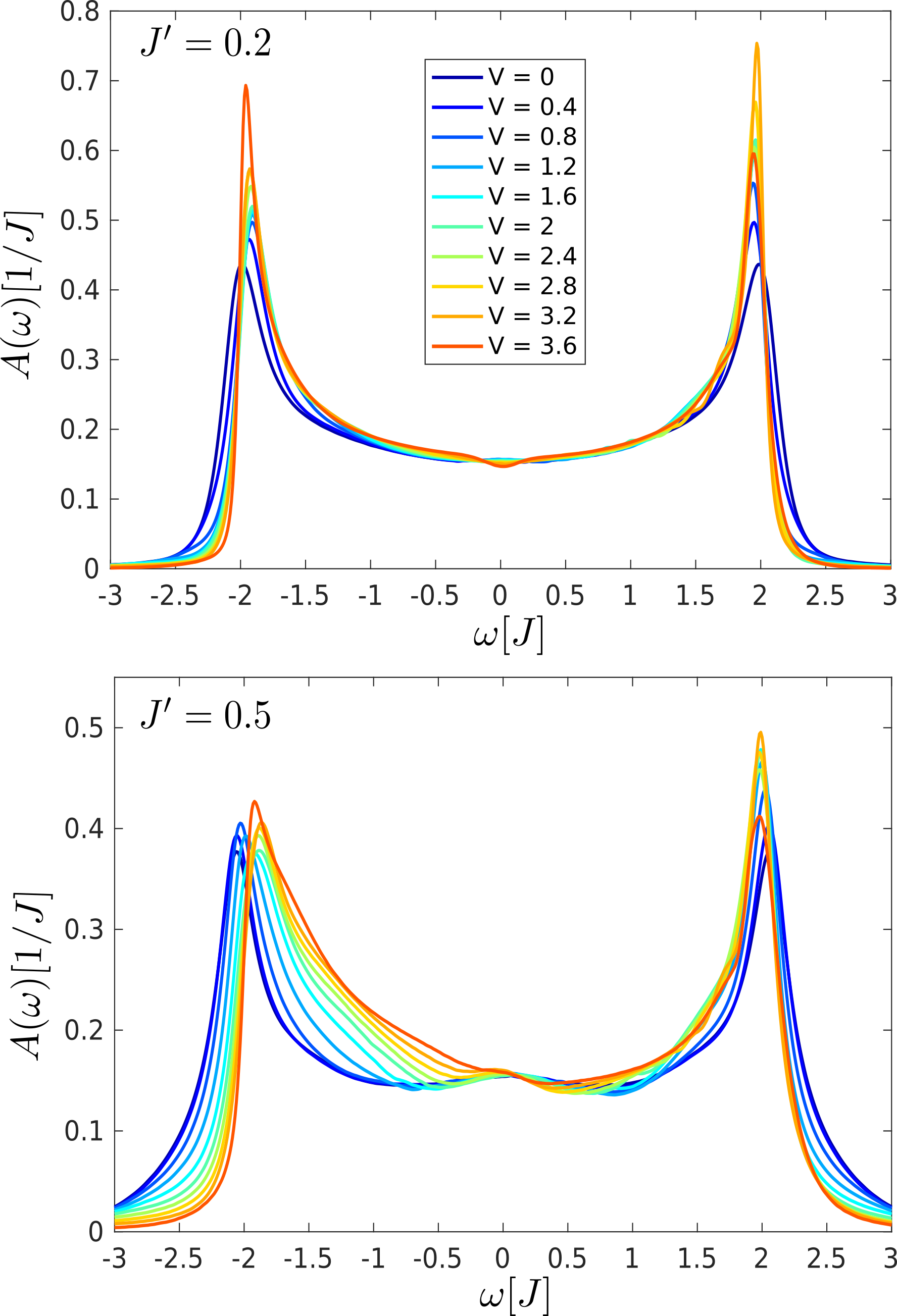}
 \caption{Local density of states for different bias voltages at site $r=-1$. Parameters are the same as in Fig.~\ref{fig:DOSimp0}.
 \label{fig:DOSImp-1}
}
\end{figure}

To make contact with the current integrands, Eq.~\ref{eq:il}, we now consider the spectral properties on the sites next to the impurity, see also Sec.~\ref{subsubsec:SSCurrent}.~\footnote{We only show results for the site $r=-1$ as the properties at $r=1$ are connected by particle-hole symmetry.}
Fig.~\ref{fig:DOSImp-1} displays the local density of states for different bias voltages. It shows two main peaks around $\omega=\pm 2$,\footnote{We note that for other values of the interaction $U$, the peaks are not fixed at $\omega=\pm U$ nor do they stay at $\omega \approx 2$ (not shown).}
and a featureless spectrum in between.
For both hybridization strengths, $J'=0.2$ and $J'=0.5$, the peaks become sharper and higher with increasing voltage. 
In addition, for $J'=0.5$ spectral weight accumulates 
 for negative frequencies up to the lower band edge at $\omega=-2$.
 
\begin{figure}
 \includegraphics[width=0.9\columnwidth]{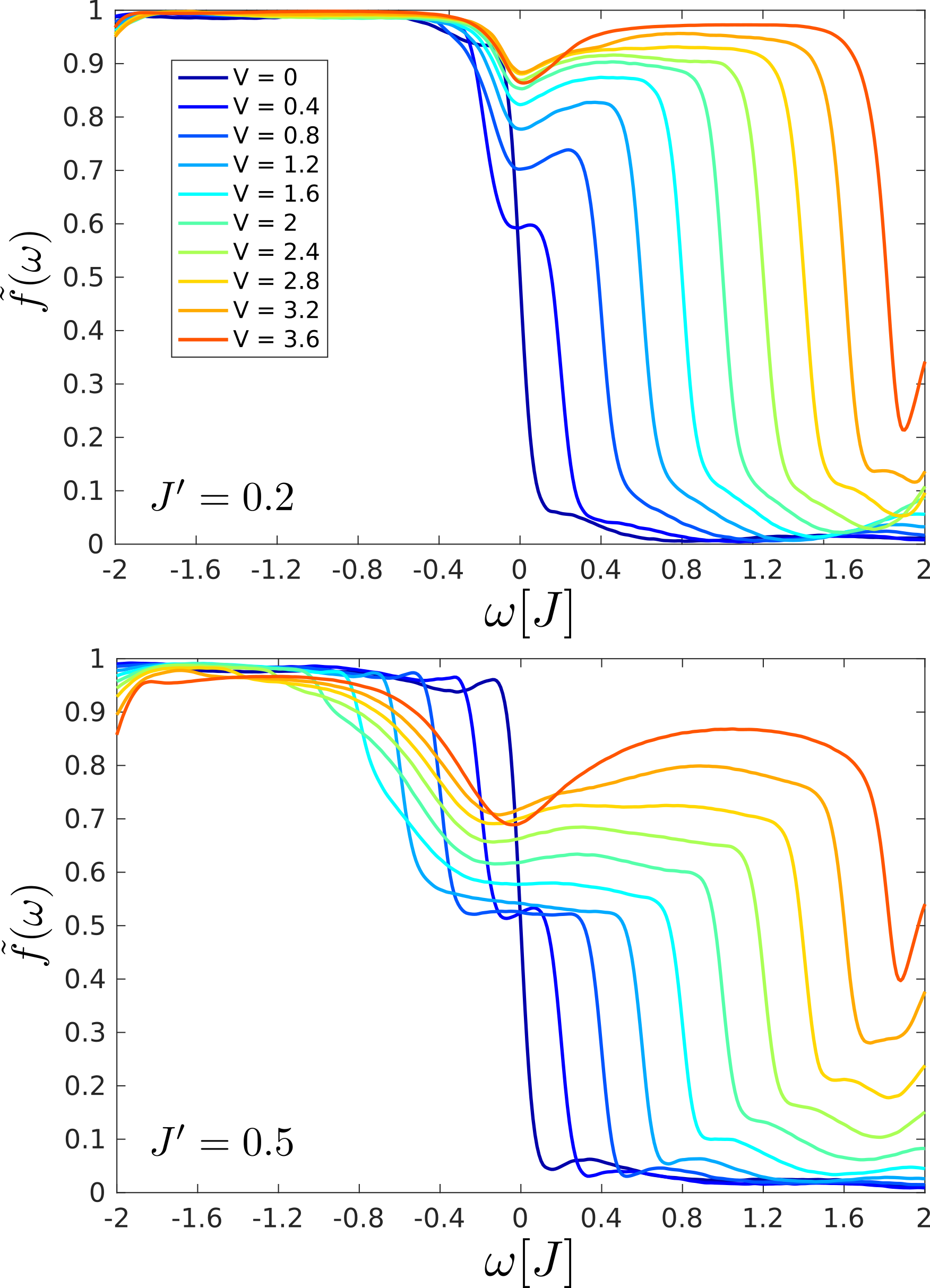}
 \caption{Same as Fig.~\ref{fig:Distimp0} but for the site $r=-1$.
 \label{fig:DistImp-1}
}
\end{figure}

A more interesting behavior can be seen in the corresponding effective distribution function for $r=-1$ presented in Fig.~\ref{fig:DistImp-1}. Similarly to the central impurity site, a double Fermi-step persists in the linear regime, while for higher bias voltages the effective distribution function becomes more similar to  the effective distribution function of the isolated left lead for which all states for frequencies smaller than its chemical potential $\mu_l$ are occupied. 
More specifically,  the plateau in the positive frequency region $0<\omega<\mu_l$ rises in the regime of the NDC. 

\begin{figure}
 \includegraphics[width=0.9\columnwidth]{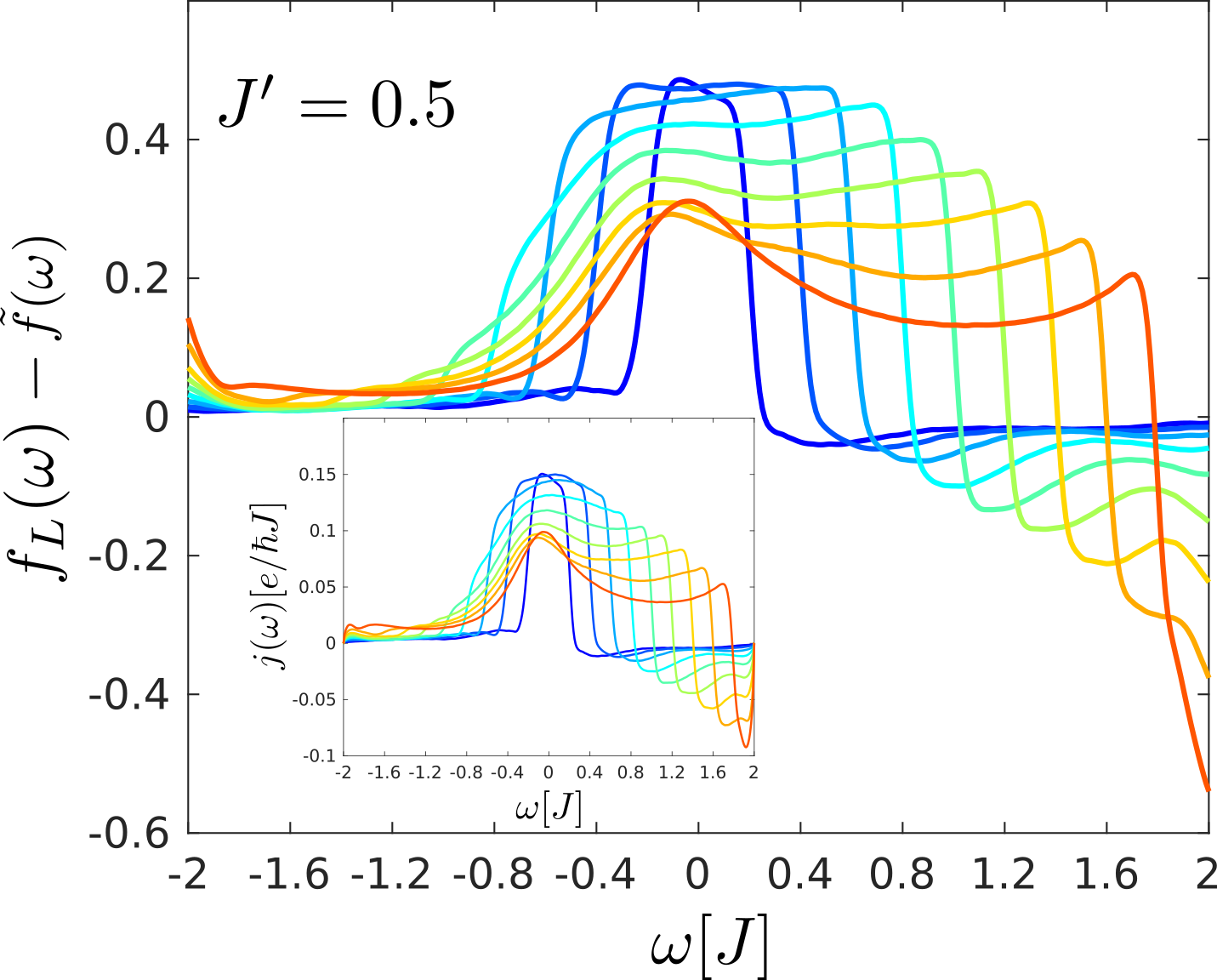}
 \caption{Difference of the effective distribution functions entering the expression for the current, Eq.~\ref{eq:il}, for different bias voltages. The inset shows the overall integrand of Eq..~\ref{eq:il}, which is dominated by the behavior of the effective distribution functions. We only present the results for $J'=0.5$. Other parameters and the label as in Fig.~\ref{fig:DOSimp0}. Note that the current integrand is identically zero outside the bandwidth, $|\omega|>2$.
 \label{fig:j_omegaimp-1}
}
\end{figure}

To elucidate
the effect of the bias dependent spectral and effective distribution functions on the current 
(cf. \eqref{eq:il}), we display in Fig.~\ref{fig:j_omegaimp-1}  for $J'=0.5$ the difference in the effective distribution functions entering Eq.\ref{eq:il} as well as the current integrand\footnote{We do not show the corresponding plots for $J'=0.2$ since they are qualitatively the same and can be readily extracted  from Fig.~\ref{fig:DOSImp-1} and Fig.~\ref{fig:DistImp-1}.} which can be seen to be dominated by the behavior of the effective distribution function.
The difference of the effective distribution functions has a Fermi-window form of amplitude $1/2$ for small voltages which is considerably distorted in the NDC regime. For negative frequencies $\omega\lesssim -1$ the amplitude quickly vanishes due to the corresponding states being filled at larger voltages, see Fig.~\ref{fig:DistImp-1} whereas at positive frequencies, the amplitude gets suppressed with increasing bias voltage which technically leads to the NDC.  Outside of the Fermi window the difference of the effective distribution functions becomes slightly negative. 
One should not overemphasize
this negative region, since the   
negative differential conductance does not depend on this\footnote{That is, the NDC is present even if these negative contributions to the current are ignored.}.

\section{Discussion and Interpretation of the results}\label{sec:Discussion}
\begin{figure}
 \includegraphics[width=0.9\columnwidth]{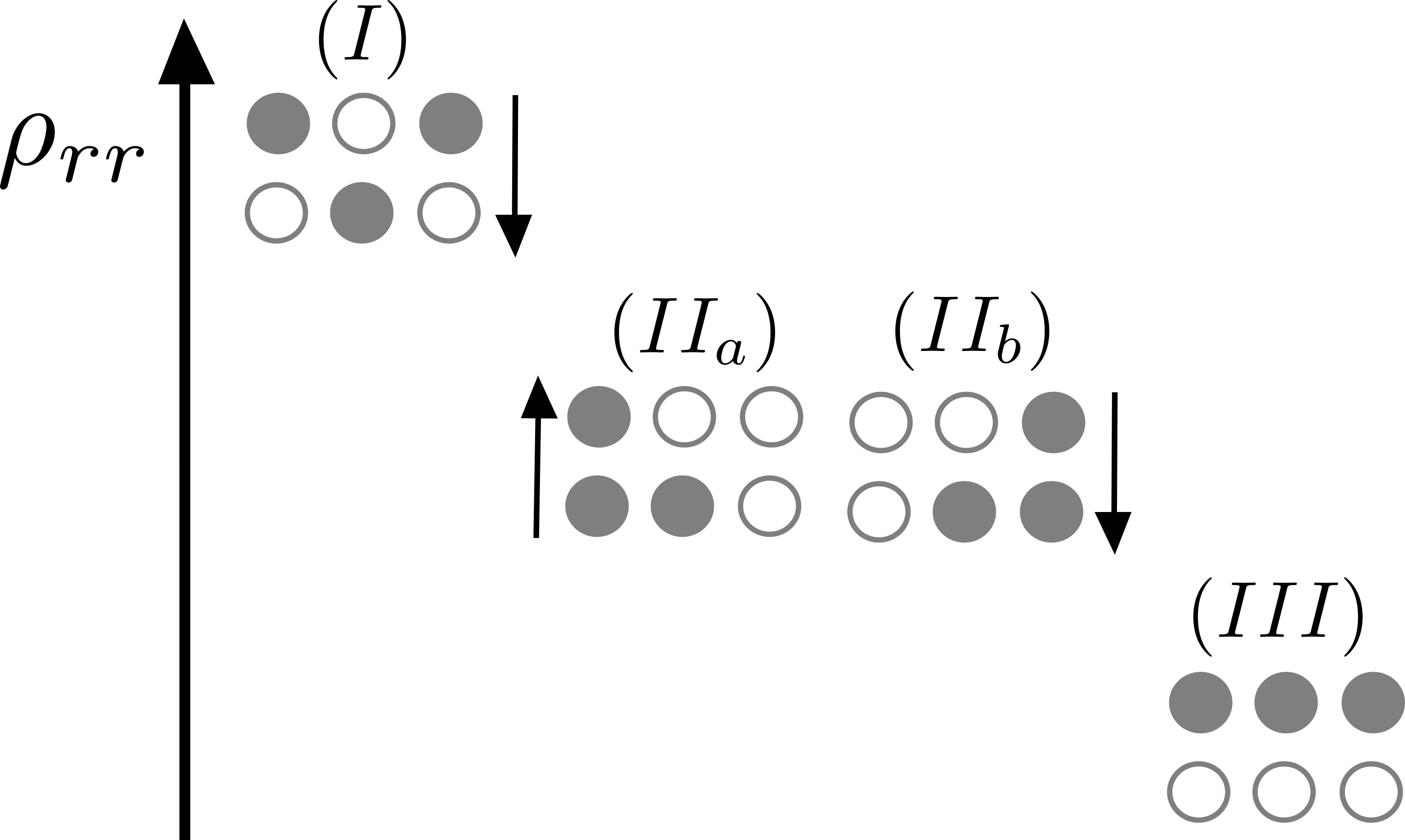}
 \caption{ Sketch of the eight different many body configurations of the interacting region, $H_{\text{dot}}$. The ordering corresponds to their respective weight in the zero-voltage case, where all states of type $(II)$ are equivalent. The arrows indicate the respective behavior for growing bias voltages in the NDC regime.
 States are displayed in PHI symmetric pairs.
 \label{fig:States}
}
\end{figure}

\begin{figure}
 \includegraphics[width=0.9\columnwidth]{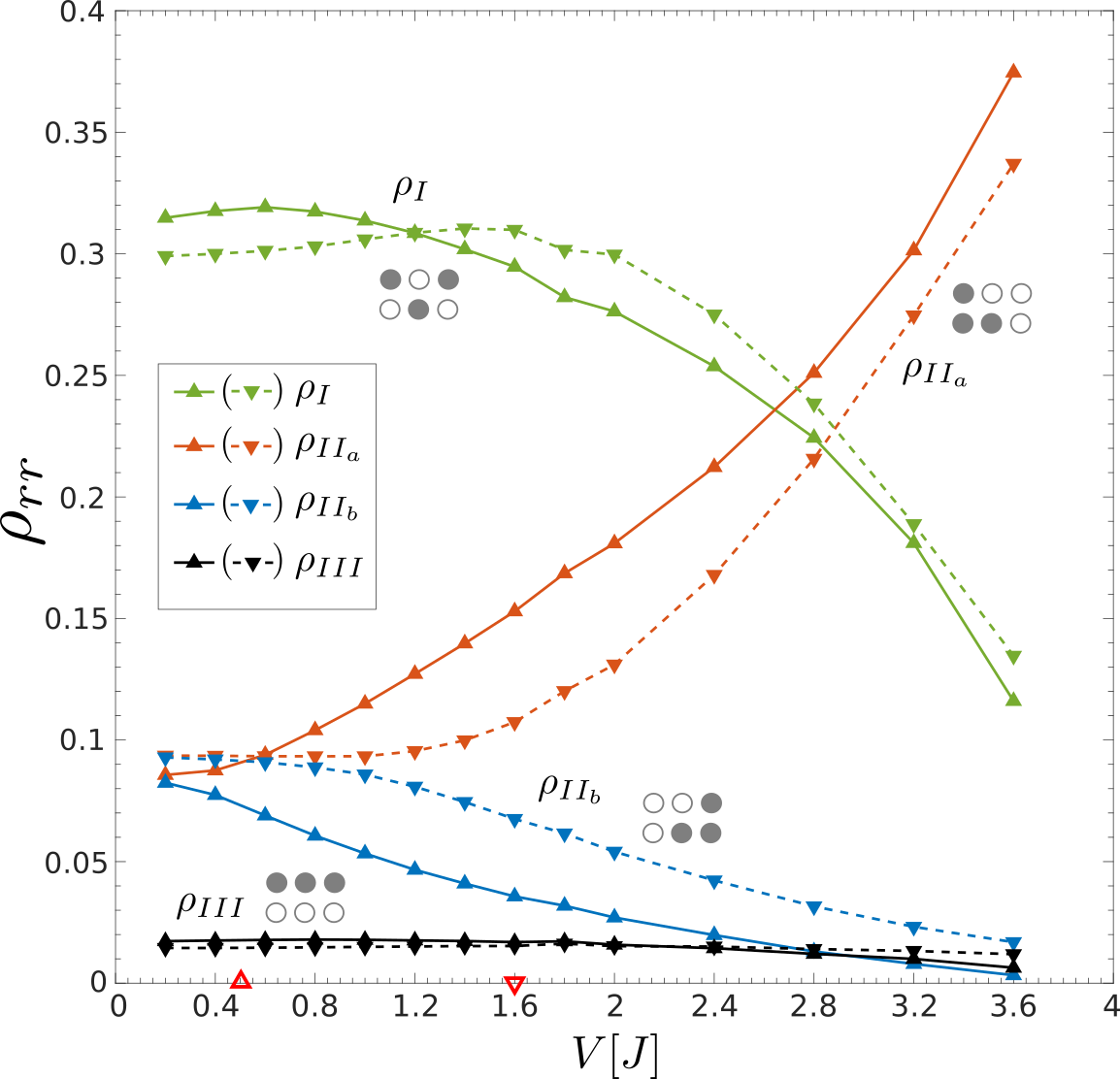}
 \caption{Probabilities of the many body configurations displayed 
in Fig.\ref{fig:States}
 for $J'=0.2$ (solid lines) and $J'=0.5$ (dashed lines). The markers on the x-axis mark the voltages corresponding to the maximum of the current, $V_{\text{max}}(J'=0.2) \approx 0.5$ and $V_{\text{max}}(J'=0.5) \approx 1.6$.
}
\label{fig:diag_rho}
\end{figure}

In order to understand the behavior of the spectral and effective distribution functions presented above, 
 we discuss the probabilities of certain characteristic many-body configuration states on the correlated sites. These are displayed in Fig.\ref{fig:States} and ranked according to their energy for zero voltage.
 Notice that the configurations in each pair are related to each other by a particle-hole+inversion (PHI) transformation~\footnote{We note that while at zero bias the model is independently particle-hole and inversion symmetric, at finite voltage the model is symmetric under a simultaneous PHI transformation.} 
and, thus, have the same probability. In addition, the states $(II_a)$ and $(II_b)$ have the same probability at zero bias voltage.
The corresponding probability is given by the 
 diagonal terms of the reduced (many-body) steady state density matrix, which is plotted
 in Fig.~\ref{fig:diag_rho} 
 as a function of the bias voltage. 

One can see that the  lowest-energy state, type $(I)$, initially slightly gains weight as the bias voltage is increased. This occurs approximately until the point where the NDC sets in. 
In the NDC regime, $V>V_{\text{max}}$, the $(I)$ state looses weight and eventually crosses with the state $(II_a)$ which becomes the dominant state at high voltages. Further, the states of type $(II_{a(b)})$, which are degenerate in equilibrium, get their degeneracy lifted by the bias voltage favoring the $(II_a)$ state since it is the one showing more occupation on the left in accordance with the chemical potentials, $\mu_L>\mu_R$. On the other hand, the weight of the highly suppressed high energy states $(III)$ stays roughly constant for all bias voltages.

\subsection{Impurity Spectral function}\label{subsec:ImpuritySpectralfunction}
As discussed above, in equilibrium ($V=0$),  the configuration $(I)$ has a large overlap with the ground state. Adding a particle at the impurity site to  $(I)$ leads to the state $(III)$. Since the energy difference, in the atomic limit $J'=0$, between these two states is $\Delta E = 2$, this process can be associated to the  $\omega\approx 2$ resonance.
The suppression of the $\omega=2$ resonance for higher voltages immediately follows from the loss of
the weight of the $(I)$ state, 
cf. Fig.~\ref{fig:diag_rho}.
It remains to explain the development of the dominant central peak for high voltages. In general, a resonance at zero frequency occurs when two low-lying states differing by one particle, at the corresponding cite, are almost degenerate.
This is the case for the states of type $(II)$. The development of the central peak is then readily explained by the increased weight of the state $(II_a)$ at high bias voltages.

\subsection{Negative differential conductance}\label{subsec:NDC}
In Sec.\ref{subsec:GF_minus1}, we discussed  that  on the level of NEGF's the NDC at large voltage in the IRLM arises due to the effective distribution function on the site next to impurity resembling the Fermi-function of the corresponding lead. 
This can be seen as
an effective decoupling of the impurity from the leads at large  bias voltage. 

Refs.~\cite{sc.dz.15,vi.sc.14} showed  that the NDC in the IRLM 
is already obtained at the Hartree-Fock (HF) level.
Therefore, it is interesting to investigate if the mechanism leading to the NDC obtained from our results is qualitatively similar to the one in the HF approximation.

\subsection{Comparison with Hartree-Fock}\label{subsubec:HF}
We will not present the details of the HF calculations, but we will only underline the connection to the AMEA results. For an alternative discussion of the NDC arising already within HF we refer to the work of Vinkler-Aviv \etal\cite{vi.sc.14}.
Within HF for the particle-hole symmetric case, which we are discussing in this paper, the Hamiltonian is the same as the non-interacting one with the only exception that 
we have a renormalized complex hopping 
between the central impurity and the $r=\pm 1$ sites:
\begin{equation}
 J' \longrightarrow \mathcal{J}_{\pm} = J' + U\langle c_{\pm 1}^\dagger c_{0}\rangle_{HF}\, .
 \label{eq:J_tilde}
\end{equation}
The computation of the GF's can be taken from the $U=0$ case, keeping in mind that the hopping ${\cal J_{\pm}}$ is complex and has to be determined self-consistently. 
It occurs that 
the local NEGF's within HF depend only on $|\mathcal{J}|^2$  and the expression for the distribution
 function on the site $r=-1$ has the form

\begin{align}
 \fs_{-1}(\omega;V) &= \frac{f_L(\omega;V) + \alpha(\omega,V)f_R(\omega;V)}{1 + \alpha(\omega,V)}\, ,
 \label{eq:HF_s}
\end{align}
where $\alpha(\omega)$ depends on the bias voltage only through $|\mathcal{J}|^2$ and  is proportional to\footnote{The precise form is $$\alpha(\omega,|\mathcal{J}|^2(V)) = |\mathcal{J}|^4 \left|\cfrac{1}{\omega^+ - |\mathcal{J}|^2 \cfrac{1}{\omega^+ - J^2 g_{TB}^R}} \frac{1}{\omega^+ - J^2 g_{TB}^R}\right|^2,$$ with $\omega^+ \equiv \omega +i0^+$ and $g_{TB}^R$ denoting the GF of a semi-infinite tight binding chain.} $|\mathcal{J}|^4$.

\begin{figure}
 \includegraphics[width=0.9\columnwidth]{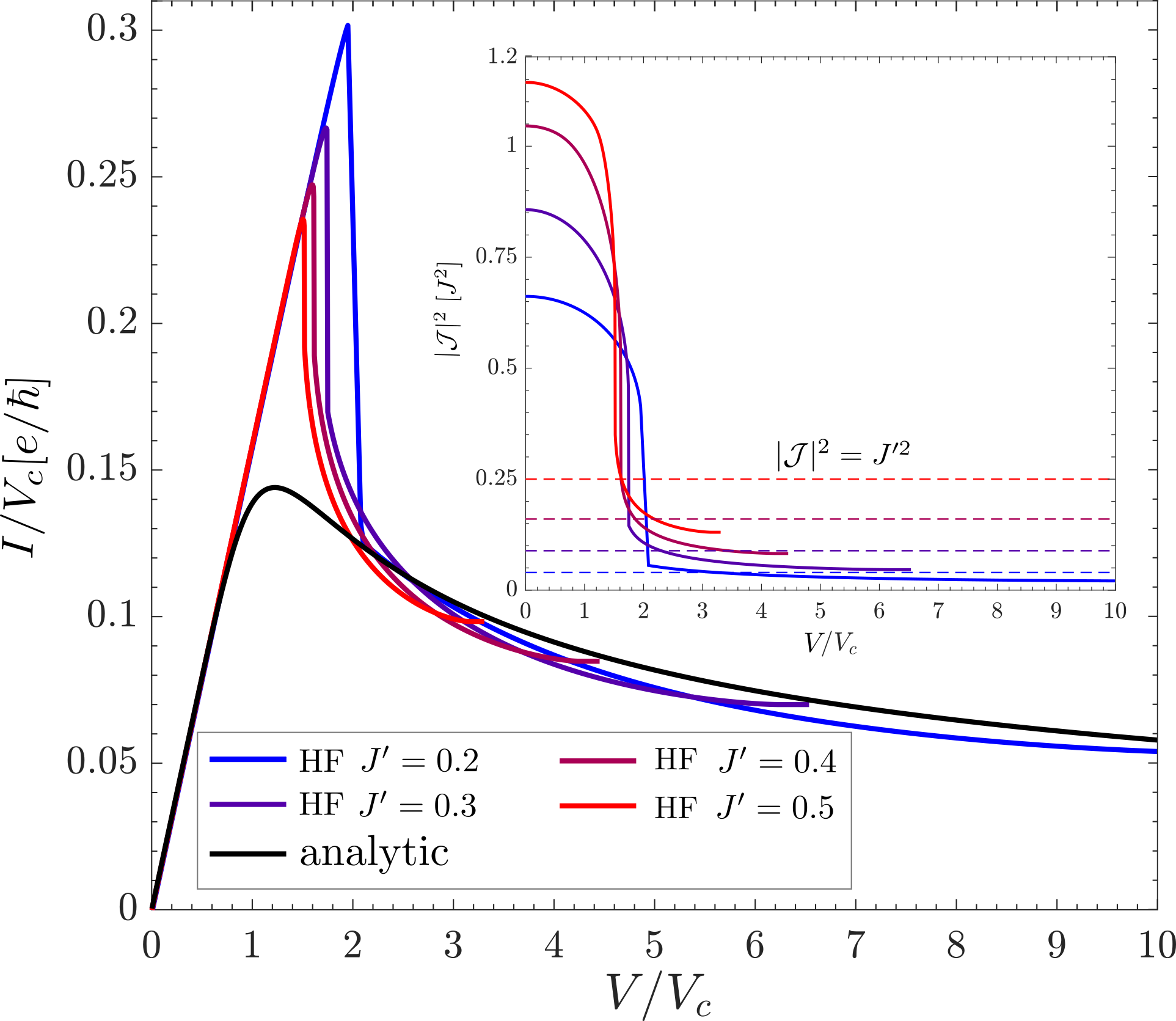}
 \caption{
Scaled current as function of scaled voltage for different hybridization strengths $0.2<J'<0.5$ (colored lines) obtained within Hartree-Fock (HF) 
 and the analytic solution, Eq.\ref{eq:Currentexpression},(black line). 
The inset shows the squared
 effective hopping amplitude $|\mathcal{J}|^2$ obtained within HF as function of the rescaled bias voltage for $U=2$, $T=0$ and different hybridization strengths $J'$. The dashed lines in the inset mark the squared bare hoppings $J'^2$.}
 \label{fig:HF_J_and_tau}
\end{figure}

In Fig.\ref{fig:HF_J_and_tau}, we display the (scaled) current and the squared effective hopping amplitude as a function of the scaled bias voltage 
within HF. The  HF current is qualitatively the same as in the exact solution but instead of a smooth transition from the linear regime to the NDC it shows a cusp and a sudden drop
\footnote{Reminiscent of a first order phase transition as is typical within a mean field treatment.} at $V/V_{c}\approx 2$. The drop in the current is accompanied by a drop in the squared effective hopping, which becomes small  for voltages outside the linear regime.
This behavior of $|\mathcal{J}|^2$ can be interpreted as an effective decoupling of the impurity from the $r=\pm 1$ sites in the NDC regime, consistent with the interpretation of the AMEA results.

In the regime in which $|\mathcal{J}|^2$ is small, i.e. large $V$, the impurity is weakly coupled to the reservoirs. Its spectral function, thus, consists of a single central peak.
 It follows that the spectral function at site $r=-1$ will be given by the DOS of a semi-infinite tight binding chain.
In addition, from Eq.\ref{eq:HF_s} it is clear that the effective distribution function $\fs_{-1}(\omega)$ will resemble the one of the left lead since $\alpha$ is strongly suppressed.
In the opposite case, when $|\mathcal{J}|^2$ is not small, $\fs_{-1}(\omega)$ will be close to a double Fermi-step and the spectral functions, independent of $r$,  will resemble the DOS of an infinite tight binding chain.
 This means that $A_{-1}^{(\text{HF})}(\omega)$ changes between two different shapes in the large and small $V$ regions in contrast to the AMEA results. 
Similar to the AMEA results, the NDC within HF is also caused by the change in the effective distribution function since the spectral density $A_{-1}^{(\text{HF})}(\omega)$ in the NDC regime has more spectral weight inside the transport window compared to the solution just before the cusp in the current.

\section{Summary and Conclusion}\label{subsec:Conclusion}
We calculated the non-equilibrium single particle Green's functions (GF), as well as the (many-body) steady state density matrix, of the Interacting Resonant Level Model (IRLM) in the presence of an applied bias voltage employing the auxiliary master equation approach (AMEA). We find developments of sidebands in the impurity spectral function which transforms into a single peak at zero energy for high bias voltages in the regime of the negative differential conductance (NDC).
Further, on the level of the non-equilibrium spectral and effective distribution functions, the negative differential conductance in the IRLM arises due to the behavior of the effective  distribution functions at the sites next to the impurity. In more detail, they feature a double fermi-step which persists in the linear regime of the current and resemble their equilibrium  form of one separated lead 
 for high bias voltages which we interpret as an effective decoupling of the system for voltages in the NDC regime. Supplementing our results with a Hartree-Fock (HF) treatment makes the decoupling explicit and shows that the spectral features resulting in the NDC are shared by both approaches.

In conclusion, our results suggest, in accordance with previous results for small interactions, an effective decoupling of the impurity from the leads as origin of the NDC in the IRLM also at the self dual point.

\begin{acknowledgments}

We would like to thank 
Delia Fugger, Irakli Titvinidze, Gerhard Dorn, and Volker Meden for fruitful discussions. A special mention goes to Antonius Dorda who originally developed most of the code used for the numeric computations.
This work was
partially supported by the Austrian Science Fund (FWF)  within 
Projects  P26508 and F41 (SFB ViCoM),  as well as NaWi Graz.
The calculations were partly performed on the D-Cluster and L-cluster Graz 
and on the VSC-3 cluster Vienna

\end{acknowledgments}

\appendix

\bibliographystyle{./prsty} 
\bibliography{./Additional_Refs,./references_database.bib}

\end{document}